\documentclass{elsarticle}
\journal{Computer Methods in Applied Mechanics and Engineering}
\usepackage{lineno,hyperref}
\usepackage{amssymb}
\usepackage{latexsym}
\usepackage{amsmath}
\usepackage{graphicx}
\usepackage{tikz}
\usepackage{color}
\usepackage{subcaption}
\usepackage{pifont}
\usepackage{natbib}
\usepackage{geometry}

\begin{document}
\begin{frontmatter}
\title{Deep Convolutional Neural Networks for Eigenvalue Problems in Mechanics}

\author[mymainaddress]{David Finol}
\author[mymainaddress]{Yan Lu}
\author[mysecondaryaddress]{Vijay Mahadevan\fnref{cor1}}
\author[mymainaddress]{Ankit Srivastava\corref{mycorrespondingauthor}}
\cortext[mycorrespondingauthor]{Corresponding author}
\ead{asriva13@iit.edu}
\fntext[cor1]{Work done prior to joining Amazon}

\address[mymainaddress]{Department of Mechanical, Materials, and Aerospace Engineering, Illinois Institute of Technology, Chicago, IL, 60616
USA}
\address[mysecondaryaddress]{Amazon AWS AI Group}

\begin{abstract}

We show that deep convolutional neural networks (CNN) can massively outperform traditional densely-connected neural networks (both deep or shallow) in predicting eigenvalue problems in mechanics. In this sense, we strike out in a new direction in mechanics computations with strongly predictive NNs whose success depends not only on architectures being deep, but also being fundamentally different from the widely-used to date. We consider a model problem: predicting the eigenvalues of 1-D and 2-D phononic crystals. For the 1-D case, the optimal CNN architecture reaches $98\%$ accuracy level on unseen data when trained with just 20,000 samples, compared to $85\%$ accuracy even with $100,000$ samples for the typical network of choice in mechanics research. We show that, with relatively high data-efficiency, CNNs have the capability to generalize well and automatically learn deep symmetry operations, easily extending to higher dimensions and our 2D case. Most importantly, we show how CNNs can naturally represent mechanical material tensors, with its convolution kernels serving as local receptive fields, which is a natural representation of mechanical response. Strategies proposed are applicable to other mechanics' problems and may, in the future, be used to sidestep cumbersome algorithms with purely data-driven approaches based upon modern deep architectures. 

\end{abstract}
\end{frontmatter}

\section{Introduction}
There has been a recent surge of interest in using deep learning using CNNs for machine learning problems in the areas of speech recognition, image and natural language processing, where advanced pattern recognition is required in data which is generally arranged in grid-like topologies \cite{rawat2017deep}. Beginning with image classification tasks, deep CNNs have consistently outperformed baseline mathematical models with prediction accuracies, in some tasks exceeding 98\% \cite{lecun1998mnist}. In several areas, deep CNNs have achieved and even surpassed human level performance. For instance, in speech recognition in particular the baseline approaches \cite{bahl1990maximum,bahl1986maximum,levinson1983introduction} have been completely outperformed by statistical learning techniques, including CNN model architectures \cite{hinton2012deep,deng2013recent}. The latter techniques have recently reached human-level accuracy \cite{Xiong2017,Stolcke2017}. This rise of the deep CNNs over the last half a decade has been aided by a similar improvement in the computational capabilities which are required by such convolutional neural networks. Graphical Processing Units have proved particularly adept at training very large CNNs using increasingly large amounts of available data\cite{schmidhuber2015deep}.

All this raises an important question: can deep architectures, such as CNNs, also be used to replace or aid certain mathematical computations central to the area of mechanics? To our knowledge, this question has not been considered before by other researchers in the field. There has admittedly been a recent push towards purely data driven mechanics computing \cite{kirchdoerfer2016data,kirchdoerfer2017data,ibanez2018manifold,mosavi2017reviewing, bessa2017framework}. However, these studies did not employ CNNs, but other predictive tools, such as regression and principal component analyses. There has also been a very recent push towards physics informed neural networks \cite{karpatne2017physics} as a way to make deep networks more data efficient, but they also seem to limit to traditional NNs or fully-connected Multilayer Perceptrons (MLPs) as their architecture. Similarly, authors have recently used deep networks for numerical quadrature calculations in FEM but have, again, limited the architecture to MLPs \cite{oishi2017computational}. More traditionally, the mechanics community has been using MLPs for various pattern recognition tasks and computations for several decades now \cite{le2015computational,Lefik2009,su2004intelligent,challis1996ultrasonic,legendre2001neural}. Furthermore, the neural networks which have been used have tended to be shallow in model depth characterized by a small number of layers, a small number of nodes per layer and a single set of computational elements. In this respect, the neural network approaches of the mechanics community resemble the pre-CNN era approaches of the machine learning community. In the machine learning community, it has been well established that deep CNN models significantly overpower both shallow and deep MLPs due to the following\cite{bengio2009learning,delalleau2011shallow,pascanu2013number,montufar2014number}:
\begin{itemize}
\item Shallow networks show poorer generalization capabilities for highly nonlinear input-output relationships.
\item Deep networks provide a more compact distributed representation of input output relationships.
\end{itemize}

A principal contribution of this paper is to develop some basic aspects of the framework which would allow deep CNNs to be employed for mechanics computations. Since no previous study exists in this domain, here we choose to focus on a simple eigenvalue problem in mechanics and train networks of various architectures to predict its eigenvalues. We show that deep CNNs massively outperform MLPs in this task. Another important contribution is to move from shallow networks in mechanics research to deep networks which are then efficiently trained over Graphical Processing Units, thereby progressing towards achieving research parity with the more traditional areas of machine learning such as vision and speech recognition. 

There are some key ideas that make CNNs particularly suitable for solving problems encountered in the mechanics field. First, CNNs employ the concept of receptive fields in their core architecture exploiting spatially local correlation in grid-like topologies \cite{goodfellow2016deep}. This implies that CNNs may represent a powerful tool for pattern recognition in various computational mechanics and materials problems which are characterized by local interactions. 

Second, there is a well-established consensus in the machine learning community that CNNs are efficiently able to learn representations which certain exhibit underlying symmetries \cite{gens2014deep,jaderberg2015spatial}. CNN architectures tend to be equivariant to general affine input transformations of the nature of translations, rotations and small distortions \cite{lecun2012learning,lenc2015understanding}. Instinctively, therefore, we expect that CNNs would be naturally suited to mechanics problems which exhibit such symmetries. These include, but is not limited to, areas in mechanics which are built on symmetries such as the mechanics and dynamics of periodic structures. 

Finally, unlike MLPs, CNNs allow for multiple channels to be associated with each input node a feature that has been successfully exploited to represent the RGB channels of individual pixels for image recognition purposes. For problems in mechanics, the concept can be naturally extended where the components of the material tensors of individual discretized elements can be identified as multiple channels to individual nodes in a CNN. CNNs require minimal preprocessing to be able to classify high-dimensional patterns from a complex decision surface \cite{lecun2015lenet}. Our aim with this paper is, then, to demonstrate the potential of CNNs in the field of computational mechanics by deploying deep CNN architectures for a narrowly focused eigenvalue problem. We compare our results with results from MLPs showing the massive performance boost that is possible with the use of CNNs. 

\section{Problem Description}

\begin{figure}[htp]
\centering
\includegraphics[scale=1]{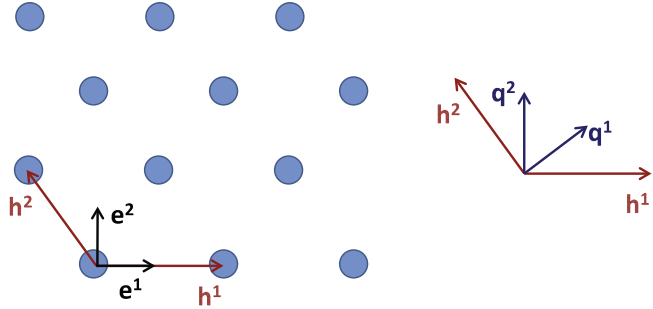} 
\caption{Schematic of a 2-dimensional periodic composite. The unit cell vectors ($\mathbf{h}^1,\mathbf{h}^2$), reciprocal basis vectors ($\mathbf{q}^1,\mathbf{q}^2$), and the orthogonal vectors ($\mathbf{e}^1,\mathbf{e}^2$) are shown.}\label{fVectors}
\end{figure}

In this section, we explain the physical model for the eigenvalue problem which we wish to emulate using CNNs. There is nothing particularly special about the chosen problem except for the fact that, being an eigenvalue problem, it represents a highly nonlinear input-output relationship and it serves well as a test bed for evaluating the capabilities of various neural networks. The frequency domain dynamics of a linear elastic medium with a spatially varying constitutive tensor $\mathbf{C}$ and density $\rho$ is given by:
\begin{equation}
\boldsymbol{\Lambda}(\mathbf{u})+\mathbf{f}=\lambda\mathbf{u};\quad \boldsymbol{\Lambda}(\mathbf{u})\equiv \frac{1}{\rho}\left[C_{ijkl}u_{k,l}\right]_{,j}
\end{equation}
where $\lambda=-\omega^2$, $\mathbf{u}$ is the displacement field, $\mathbf{f}$ is the body force, and $\boldsymbol{\Lambda}$ is a linear differential operator. For phononic problems the problem domain is periodic and is defined by a repeating unit cell. For the general 3-dimensional case, the unit cell ($\Omega$) is characterized by 3 base vectors $\mathbf{h}^i$, $i=1,2,3$. Any point within the unit cell can be uniquely specified by the vector $\mathbf{x}=H_i\mathbf{h}^i=x_i\mathbf{e}^i$ where $\mathbf{e}^i$ are the orthogonal unit vectors and $0\leq H_i\leq 1,\forall i$. The unit cell is associated with a set of reciprocal base vectors $\mathbf{q}^i$ such that $\mathbf{q}^i\cdot\mathbf{h}^j=2\pi\delta_{ij}$. Reciprocal lattice vectors are represented as a linear combination of the reciprocal base vectors, $\mathbf{G}^\mathbf{n}=n_i\mathbf{q}^i$, where $n_i$ are integers. Fig. (\ref{fVectors}) shows the schematic of a 2-D unit cell, clearly indicating the unit cell basis vectors, the reciprocal basis vectors and the orthogonal basis vectors.
The material properties have the following periodicity:
\begin{equation}
C_{jkmn}(\mathbf{x}+n_i\mathbf{h}^i)=C_{jkmn}(\mathbf{x});\quad \rho(\mathbf{x}+n_i\mathbf{h}^i)=\rho(\mathbf{x})
\end{equation}
where $n_i(i=1,2,3)$ are integers. Due to the periodic nature of the problem, it accepts solutions of the form $\mathbf{u}(\mathbf{x})=\mathbf{u}^p(\mathbf{x})e^{\mathrm{i}\mathbf{k}.\mathbf{x}}$ where $\mathbf{k}$ is the Bloch wavevector with components $\mathbf{k}=Q_i\mathbf{q}^i$ where $0\leq Q_i\leq 1,\forall i$ and $\mathbf{u}^p$ is a periodic function. Under the substitution, the harmonic elastodynamic problem can be formally written as (neglecting the body force):
\begin{equation}
\boldsymbol{\Lambda}^{(\mathbf{k})}(\mathbf{u}^p)=\lambda\mathbf{u}^p
\label{eq:Eigenvalue}
\end{equation}
where the superscript $(\mathbf{k})$ is now included to emphasize that the operator depends upon the Bloch wavevector. Explicitly we have:
\begin{equation}
\boldsymbol{\Lambda}^{(\mathbf{k})}(\mathbf{u}^p)\equiv \frac{1}{\rho}\left[C_{ijkl}u_{k,l}^p\right]_{,j}+\frac{\mathrm{i}q_jC_{ijkl}}{\rho}u_{k,l}^p+\frac{\mathrm{i}q_l}{\rho}\left[C_{ijkl}u_k^p\right]_{,j}-\frac{q_lq_jC_{ijkl}}{\rho}u_k^p
\end{equation}
For a suitable span of the wavevector, the sets of eigenvalues $\lambda_n$ (and the corresponding frequencies $\omega_n$) constitute the phononic dispersion relation of the composite. There are several numerical techniques for calculating the eigenvalues but a common method is to expand the field variable $\mathbf{u}^p$ in an appropriate basis and then use the basis to convert the differential equation into a set of linear equations. Both the Plane Wave Expansion\cite{kushwaha1993acoustic} method and Rayleigh quotient\cite{lu2016variational} follow this strategy.

\subsection{The 1-D Problem}
\begin{figure}[htp]
\centering
\includegraphics[scale=.7]{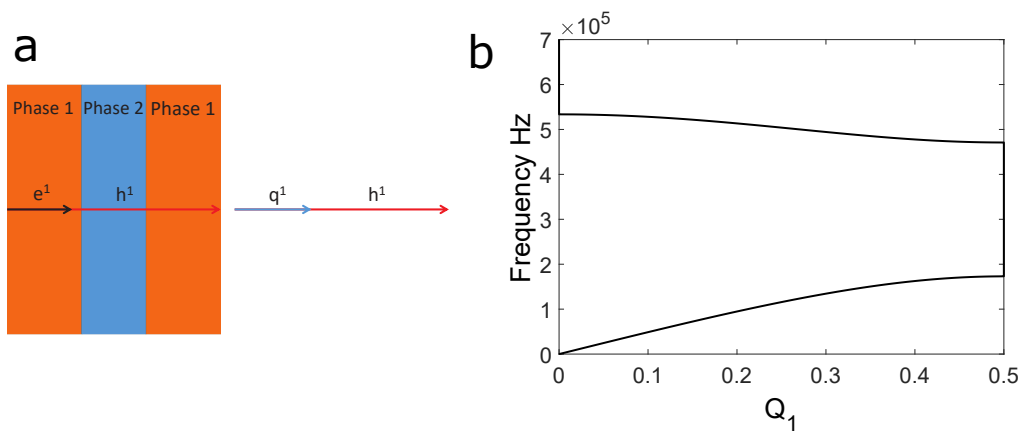}
\caption{a. Schematic of a 1-dimensional 2-phased periodic composite. The unit cell vector ($\mathbf{h}^1)$, reciprocal basis vector ($\mathbf{q}^1$), and the orthogonal vector ($\mathbf{e}^1$) are shown, b. Bandstructure of a 1-D, 2-phase phononic crystal showing the frequency eigenvalues constituting the first two passbands. Unit cell details in \cite{srivastava2016metamaterial}.}\label{1-DResult}
\end{figure}

There is only one possible Bravais lattice in 1-dimension with a unit cell vector whose length equals the length of the unit cell itself (Fig. \ref{1-DResult}a). Without any loss of generality we take the direction of this vector to be the same as $\mathbf{e}^1$. If the length of the unit cell is $a$, then we have $\mathbf{h}^1=a\mathbf{e}^1$. The reciprocal vector is given by $\mathbf{q}^1=(2\pi/a)\mathbf{e}^1$. The wave-vector of a Bloch wave traveling in this composite is specified as $\mathbf{k}=Q_1\mathbf{q}^1$. To completely characterize the band-structure of the unit cell it is sufficient to evaluate the dispersion relation in the irreducible Brillouin zone ($0\leq Q_1\leq .5$). 
For plane longitudinal wave propagating in the $\mathbf{e}^1$ direction the only displacement component of interest is $u_1$ and the only relevant stress component is $\sigma_{11}$. The equation of motion and the constitutive law are:
\begin{equation}\label{equationofmotion1D}
\sigma_{11,1}=-\lambda\rho(x_1) u_1; \quad \sigma_{11}=E(x_1)u_{1,1}
\end{equation}
where $E(x_1)$ is the spatially varying Young's modulus. The exact dispersion relation for 1-D longitudinal wave propagation in a periodic layered composite can be solved using the transfer matrix method \cite{srivastava2016metamaterial}. For 2-phase composites, the relation is particulary simple \cite{rytov1956acoustical},
\begin{eqnarray}\label{Rytov}
\nonumber \cos(\mathbf{k} a)=\cos(\omega h_1/c_1)\cos(\omega h_2/c_2)-\Gamma \sin(\omega h_1/c_1)\sin(\omega h_2/c_2),\\
\Gamma=(1+\kappa^2)/(2\kappa), \; \kappa=\rho_1c_1/(\rho_2c_2), 
\end{eqnarray}
where $h_i$ is the thickness, $\rho_i$ is the density, and $c_i$ is the longitudinal wave velocity of the $i$th layer $(i = 1,2)$ in a unit cell. We can solve for the corresponding wave number $\mathbf{k}$ (or, equivalently, $Q_1$) by providing a frequency, $\omega$ (or, equivalently, $f=\omega/2\pi$), using (\ref{Rytov}). These $f-Q_1$ (or $\omega-\mathbf{k}$) pairs constitute the eigenvalue band-structure of the composite when the wavevector is made to span the first Brillouin zone ($Q_1\in[0,0.5]$). A representative example of the bandstructure, calculated for a representative 1-D, 2-phase composite (Fig. \ref{1-DResult}a), is shown in Fig. (\ref{1-DResult}b). The frequency values in Fig. (\ref{1-DResult}b) are, thus, related to the eigenvalues of the phononic crystal when a certain wavenumber $Q_1$ is specified. Results in Fig. (\ref{1-DResult}) are calculated by using the exact physical solution of the system given by the Rytov equation (\ref{Rytov}). Our aim in this paper is to train neural networks of varying architectures to sidestep the physical model as represented by Eqs. (\ref{equationofmotion1D},\ref{Rytov}). 

\subsection{Input-output relationship framework}
The phononic eigenvalue problem described above can be represented as an input-output relationship. Formally, we can write:
\begin{equation}
\mathbf{e}=\mathbf{f}_c(E(x_1),\rho(x_1),Q_1)
\label{eq:1DIOC}
\end{equation}
where $\mathbf{e}$ is a vector of eigenvalues which is obtained through a vector of nonlinear functions $\mathbf{f}_c$ which operates on the space dependent material properties ($E,\rho$) and a choice of the wavenumber $Q_1$. An approximation to the eigenvalues can be generated by considering the material properties as discretely defined over the unit cell. We first normalize the unit cell with its length and discretize the range of $x_1$ ($x_1\in [0,1]$) into $N$ segments. We identify the material properties over these segments as the input variables. The material properties now become vectors themselves, individually defined over each segment, and the input-output relationship becomes:
\begin{eqnarray}
\mathbf{e}=\mathbf{f}_d(\mathbf{E},\boldsymbol{\rho},Q_1),\; \mathbf{E}=\{E_i\},\boldsymbol{\rho}=\{\rho_i\};\;i=1,2...N
\label{eq:1DIOD}
\end{eqnarray}
where $\mathbf{f}_d$ are the sought approximations to the continuous input-output relationships $\mathbf{f}_c$ in (\ref{eq:1DIOC}). In this paper we have taken $N=100$ and we have sought to train neural networks to learn and predict $\mathbf{f}_d$ from a set of training data. The training data consists of input and output datasets created from the solution of the exact problem (\ref{equationofmotion1D},\ref{Rytov},\ref{eq:1DIOC}). Specifically, the input data consists of randomly generated instantiations of vectors $\mathbf{E},\boldsymbol{\rho}$ and the output data consists of the corresponding first two eigenvalues, appropriately normalized, at chosen values of $Q_1$. For the set of problems under consideration, we seek to estimate the first two eigenvalues at 10 different $Q_1$ points within the range of $Q_1$ ([0,0.5]). This essentially translates into an output vector size of 20 and a flat input vector size of 200. 100 elements of the input vector correspond to $\mathbf{E}$ and the rest correspond to $\boldsymbol{\rho}$. In training neural networks to learn this input-output relationships, there are some questions of primary concern. Some of these pertain to the architecture of the network, the number of training data needed to achieve a certain desired error, the method of training the network, the method by which the training data is generated, the effect of the training data on the ability of the network to generalize for unseen examples, and in unseen regions of the space. These are discussed in detail in the next sections. 

\section{Neural Networks and the Representation Learning Approach}

\begin{figure}[htp]
\centering
\includegraphics[scale=0.7]{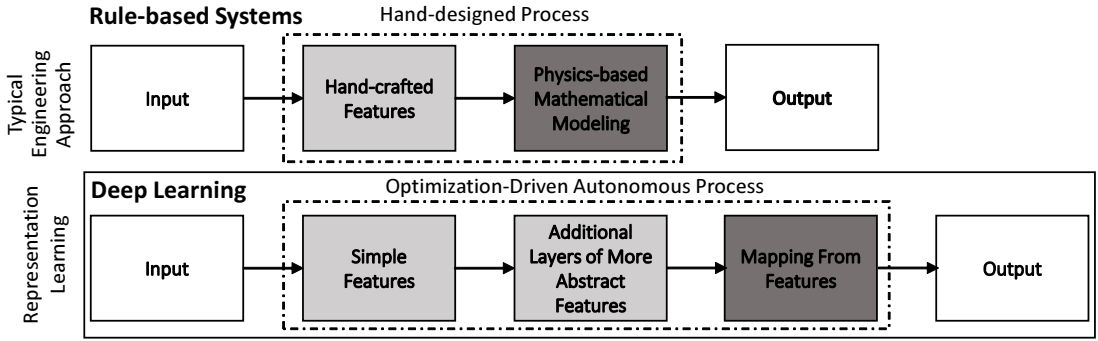}
\caption{Manual feature crafting vs. the representation learning approach}
\label{fig:replearning}
\end{figure}

A central problem in mechanics, as in other areas, is the approximation of a function which relates an input to an output in a system of interest (\ref{eq:1DIOC}). The traditional method of attempting this problem is to manually convert the inputs to a set of representative features, which could then be related to the outputs. Representation learning, on the other hand, is a set of techniques that allows a system to automatically discover the optimal representations needed for feature detection, classification or real-value mapping from raw data. This replaces manual feature engineering and allows a machine to both learn the features and use them to perform a specific task (Fig. \ref{fig:replearning}). Neural networks are a set of algorithms which perform this task by creating automatic representations of input data in their hidden layers. The first precursors to the modern neural networks were proposed by Rosenblatt \cite{rosenblatt1958perceptron}. Since then, significant advances in the area of machine learning has led to modern neural networks with complex function approximation capabilities \cite{schmidhuber2015deep}. Artificial Neural Networks (ANN) are comprised of multiple interconnected simple computational elements called \emph{neurons}. Both the fully-connected multi-layer perceptrons and convolutional neural networks fit within the class of ANNs and they differ primarily in their neuronal architecture and interconnectivity. 

\subsection{Multi-Layer Perceptrons}
\begin{figure}[htp]
\centering
\includegraphics[scale=1]{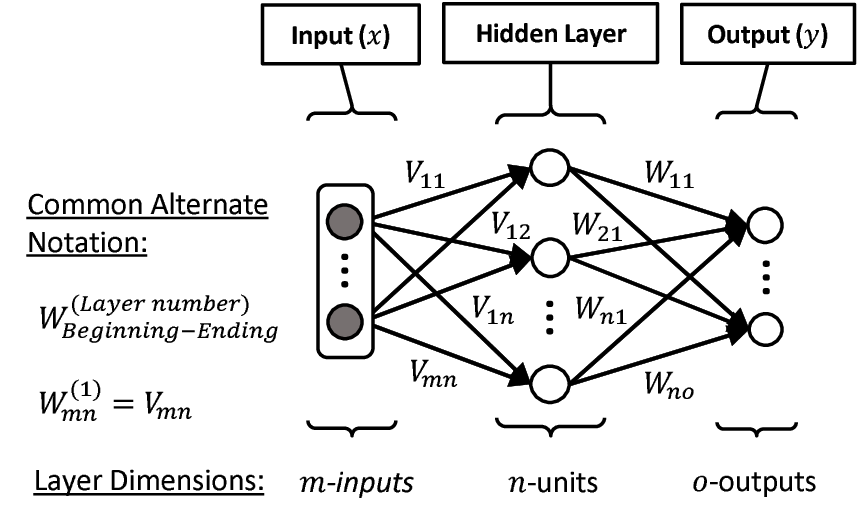}
\caption{Graph of a simple Multi-layer Perceptron with fully-connected layers}
\label{fig:FCMLP}
\end{figure}

The Multi-layer Perceptron (MLP) concept, more generally referred as feed-forward neural network, was initially proposed as an universal function approximator \cite{hornik1991approximation}.  The objective of an MLP is to approximate a function $f$ between inputs $\mathbf{x}$ and outputs $\mathbf{y}$: 
\begin{equation}{}
\mathbf{y} = f(\mathbf{x})
\end{equation}
This process can be represented in NN terms as a function mapping of inputs to the outputs through a set of optimizable parameters. For the single hidden layer MLP shown in Fig. (\ref{fig:FCMLP}), these optimization parameters are the weight matrices $\mathbf{V,W}$ and the mapping is approximated as:
\begin{equation}{}
y \approx f(x,\theta)= \mathbf{W}g(\mathbf{V}x)
\end{equation}
where appropriate matrix multiplications are assumed. $g$ represents the activation function which is generally a non-linear transformation. The weights of the network are typically stochastically initialized and are subsequently tuned within the training phase of the network. One method of training the network is to relate its known input-output data and minimize its prediction deviation from the output by appropriately changing the weights, through an optimization process.
%
%
%
%

\subsection{Convolutional Neural Networks}
 Three key ideas differentiate convolutional networks from conventional networks, which have made them highly successful in various engineering and science fields \cite{Ji2013,Sermanet2013,bojarski2016end,aurisano2016convolutional,wallach2015atomnet,tajbakhsh2016convolutional,goh2017deep} : the use of convolution operation, the implementation of the Rectified Linear (ReLU) activation function, and a representation-invariance imposing operation termed Pooling. 
 
\subsubsection{Convolution} 
An important difference between a fully-connected MLP and a CNN is the use of the convolution operation instead of the standard matrix multiplication operation. While in the case of the MLP in Fig. (\ref{fig:FCMLP}) the inputs to the neurons in the middle layer are obtained by a simple matrix multiplication of the weight $\mathbf{V}$ and the input $\mathbf{x}$, for a CNN, the input will instead be transformed using a convolution operation through a kernel $\mathbf{w}$ into a feature map $\mathbf{h}$. Assuming that the input is a 2-D vector of length $N$ and depth $d$, this can be represented as a 3-D matrix of size $N\times 1\times d$. The kernel is similarly assumed to be a matrix of size $m\times 1\times d$ where $m<N$. In this case, one element of the feature map will be calculated by computing the Einstein sum $h_l=x_{j+l,1,k}w_{j,1,k};j=1,...m,k=1,...d$ where $l$ is the current location of the filter. The filter is then advanced from its current location by a predetermined step size termed stride and the next element of the feature map is calculated. In our examples presented below, a good value of the stride is determined to be 1. This process is repeated until the kernel spans $\mathbf{x}$ in the length dimension and completes the calculation of $\mathbf{h}$. This process is symbolically represented as:
\begin{equation}{}
\mathbf{h} = \mathbf{x}*\mathbf{w}
\end{equation}
where $*$ represents the convolution operation. For instance, in our 1D phononic eigenvalue problem, the inputs are the spatially ordered sequence of material property values. These are represented by an input vector $\mathbf{X}$ whose depth is 2. In the depthwise direction, the first element corresponds to the modulus of a given finite element and the second element corresponds to the density value. The convolutional layer, for our case, convolves each input matrix with $k$ kernels ${\mathbf{w}_i}$ resulting in a total of $k$ feature maps. Each feature map $\mathbf{h}_i$ is computed as follows: 
\begin{equation}{}
\mathbf{h}_i = \mathbf{X}*\mathbf{w}_i + b_{i},i = 1,...,k. 
\end{equation}
where ${b_{i}}$ are bias parameters. CNNs work by tuning the parameters of the kernels $\mathbf{w}_i$ and the bias parameters $b_i$ in order to learn the desired input-output relationships. There are some interesting points to note here. First, the feature maps corresponding to the modulus and density channels are not completely independent of each other as they are generated from the dot product of the same kernel $\mathbf{w}_i$. This allows the neural network to associate the different material properties at a point as belonging to the same material. Second, the concepts of kernels allows for the feature maps to represent local interactions. This sparsity allows CNNs to learn an input-output relationship dependent upon spatial and temporal structure more efficiently than MLPs.

\subsubsection{ReLU Activation} 
Having calculated the pre-activation feature maps $\mathbf{h}_i$, they are passed through nonlinear activation functions commonly referred to as ReLU. The Rectifier Linear Unit Rectifier Linear Unit (ReLU) \cite{Glorot2011} has proven to provide high computational efficiency and is a piece-wise linear activation function that outputs zero if the input is negative and outputs the input if it is greater than zero. Mathematically, given a feature map $\mathbf{h}_i$, a ReLU function is defined as follows: 
\begin{equation}{}
\hat{\mathbf{h}}_ {i} = \mathrm{max}(0, \mathbf{h}_{i})  
\end{equation}
in which $\mathbf{h}_i$ and $\hat{\mathbf{h}}_i$ represent the ReLU input and output respectively. 

\subsubsection{Pooling Operation}
 
\begin{figure}[htp]
\centering
\includegraphics[scale=0.4]{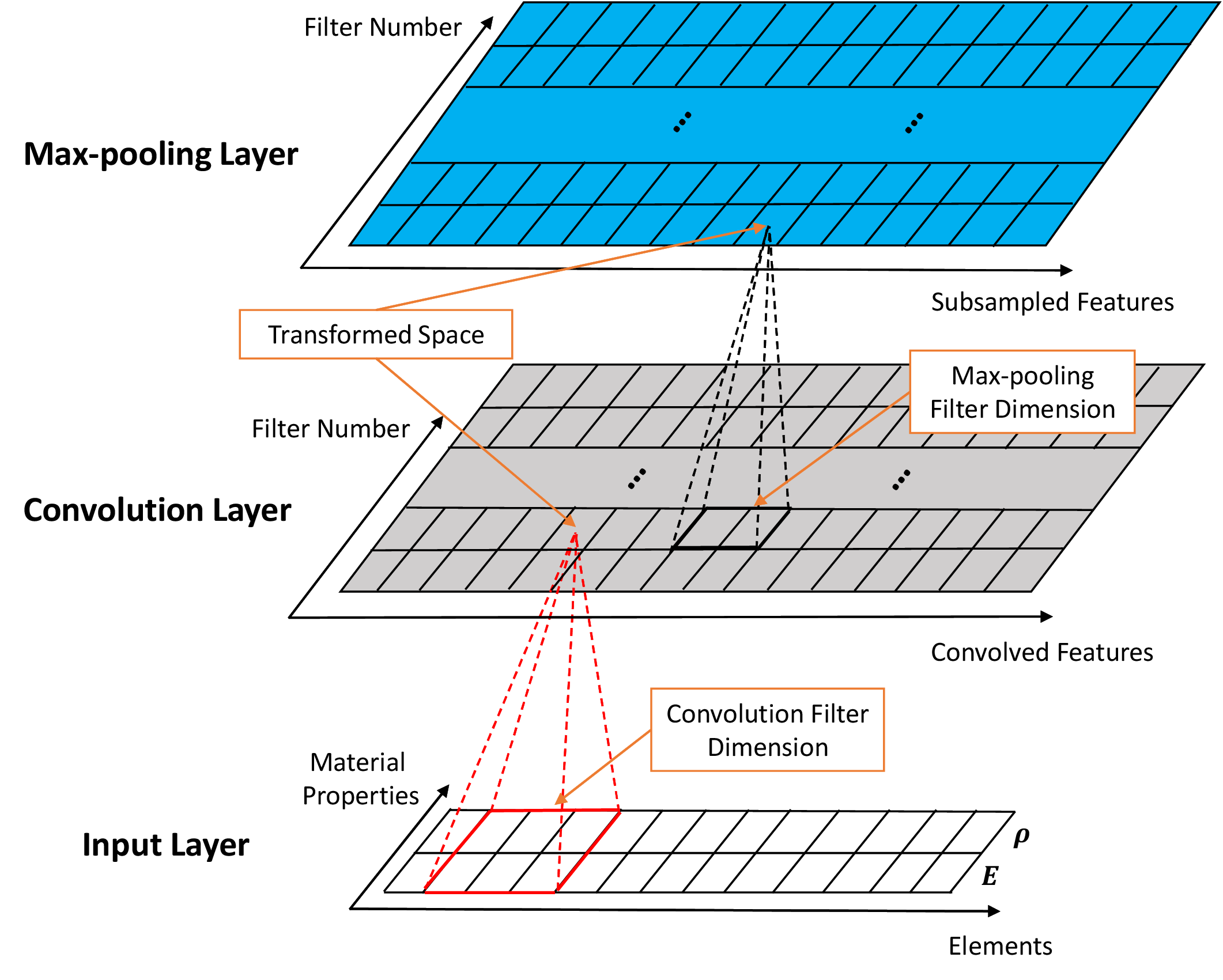}
\caption{Convolution and max-pooling transformations of modulus and density axes}
\label{fig:convMax}
\end{figure}

Once the convolution operations have been applied along with ReLU activations, the outputs pass through a parameter-reducing layer commonly referred to as a max-pooling layer. The output of this layer is the maximum unit from $p$ adjacent input units. In our 1-D problem, pooling is performed along both the transformed modulus and density data axes (Fig. \ref{fig:convMax}.)  Following the findings of optimal CNN achitectures in computer vision and speech recognition \cite{LeCun1995,krizhevsky2012imagenet} and our own optimization work, the pooling operation is performed only once in our case, after two consecutive convolution layers. The general intuition is that as more pooling layers are applied, units in higher layers would be less discriminative with respect to the variations in input features \cite{Zhang2017}. 

\subsection{Optimization Algorithm}

For both MLPs and CNNs in this paper, we employed the Stochastic Gradient Descent with Momemtum (SGDM) through backpropagation for network training.  The training is performed on small groups of datasets at a time, termed mini-batches. The algorithm updates the model parameter by taking small steps in the direction of negative gradient:
\begin{equation}{}
\theta_{l+1} = \theta_l - \alpha \nabla E(\theta_l)+\gamma(\theta_{l}-\theta_{l-1}) 
\end{equation}
where $\theta_l$ represents the optimizable model parameters at iteration $l$,  $\alpha$ is the learning rate, $E(\theta_l)$ is the current minibatch cost function and the $\gamma(\theta_{l}-\theta_{l-1})$ is the contribution of the previous gradient step to the current iteration \cite{bishop2006pattern}. The specific algorithm employed is termed as Adam and it involves adaptive learning rates for different parameters from estimates of first and second moments of the gradients \cite{Kingma2014}.

\subsection{Objective Function} 
 
Hung et al and Lippmann \cite{Hung1996,Richard1991} have demonstrated that neural networks based on squared-error functions are able to accurately estimate posterior probabilities for finite sample sizes. The mean squared error objective function is defined as, 
\begin{equation}{}
MSE =  \frac{1}{m} \sum {(\hat{y} -y)}^2
\end{equation}
where $\hat{y}$ is the prediction and $y$ is the actual target output (eigenvalue in our case.) The summation is over all outputs and over all the data points in a mini-batch and $m$ is the total number of terms in the summation. 

\subsection{Data Mapping and Normalization}

As is standard practice in modern neural network implementations, all input material property data is normalized through mean normalization:
\begin{equation}{}
x_{norm} = \frac{x-\bar{x}}{x_\mathrm{max}-x_\mathrm{min}} 
\end{equation}
where $\bar{x}$ is the mean of the input vector. Furthermore, the eigenvalues are also normalized by a reference maximum, as is standard practice in the machine learning field.

\section{Numerical Experiments}

\subsection{Data Generation}

The input data for the CNN described above was generated following standard physical principles that guide the phononic eigenvalue problem as previously described. The training, validation, and test data sets can be divided into two broad categories. In the first category, the material property vectors $\mathbf{E}$ and $\boldsymbol{\rho}$ were generated using a uniform probability distribution on the modulus and density values. The probability distribution was created with lower and upper bounds for both the modulus (lower bound: 100MPa, upper bound: 300GPa) and density (lower bound: 800kg/m$^3$, upper bound: 8000kg/m$^3$). From these distributions, we created 100-element unit cells $\Omega^i$ randomly generated , which are characterized by modulus vectors $\mathbf{E}^i$ and density vectors $\boldsymbol{\rho}^i$. Corresponding to these unit cells, we calculated the 20 eigenvalues of interest which constitute the output data vectors $\mathbf{e}^i$. This dataset category is termed Dataset-A for future reference and has a total of 300,000 data samples. In the respective results, a small fraction of this dataset was employed purely for model training, validation and, in an initial phase, testing of the networks. We note that since the modulus and density values in Dataset-A are independently sampled from two different probability distributions, the material properties assigned to the individual elements in any given $\Omega^i$ very likely do not correspond to any real material.

The second broad category of data used in this paper is aimed at testing specific capabilities of the model in making predictions for those unit cells $\Omega$ which might be of more practical interest but which are highly unlikely to be represented in Dataset-A. One such example is the case where $\Omega$ is composed of only 2 different material phases - a configuration which appears frequently in the phononics and metamaterial literature \cite{hussein2014dynamics,srivastava2016metamaterial,srivastava2017evanescent}. To create a dataset corresponding to this configuration, we divided the 100-element unit cell into two zones of 50 elements each. All elements in these individual zones were then assigned randomly generated but same modulus and density values. The input-output data for this dataset is referred to as Dataset-2. We similarly created datasets for 3 and 10 material phases and refer to them as Dataset-3 and Dataset-10 respectively which, in addition to Dataset-2, is collectively referred to as Dataset-B. At this point it should be noted that it is highly unlikely that any individual case appearing in Dataset-B also appears in Dataset-A.

\subsection{Model Architectures}

\begin{figure}[htp]\label{fig:cnn2new}
\centering 
\includegraphics[scale=0.77]{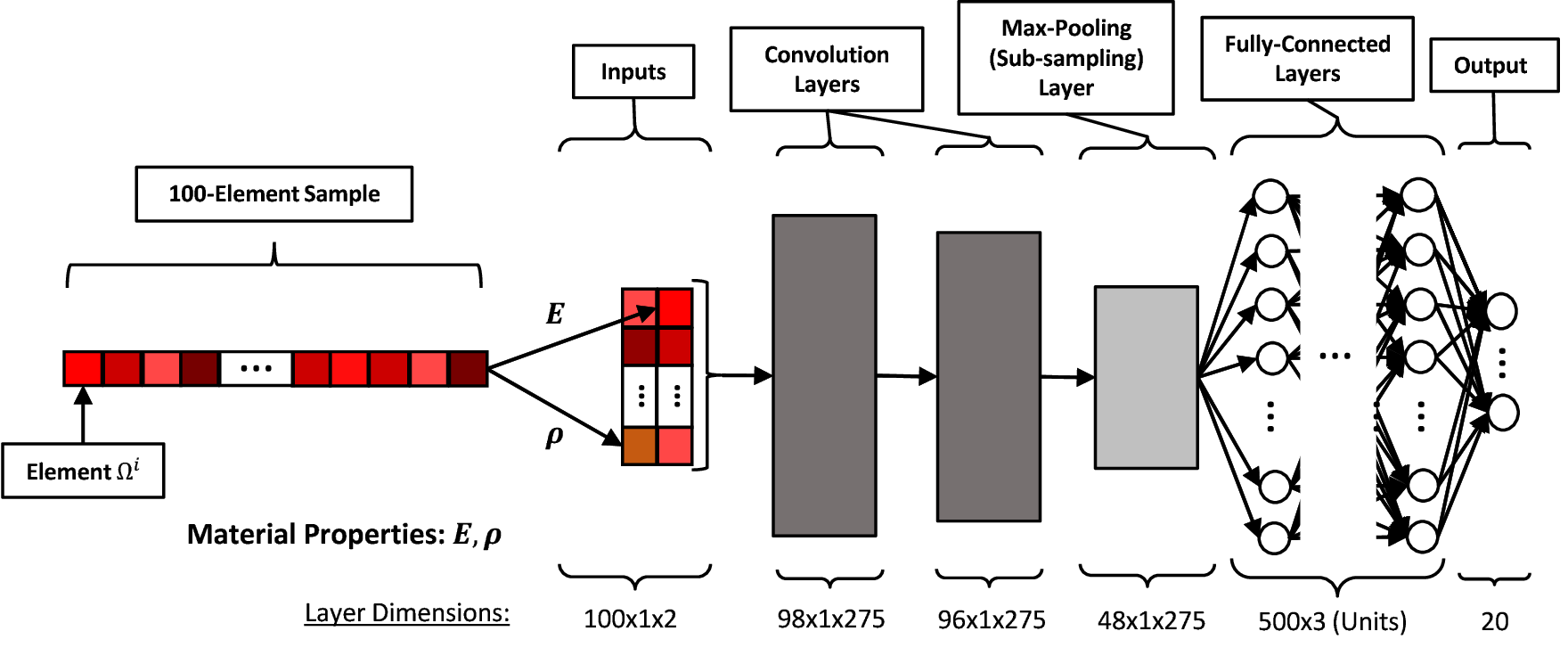}
\caption{CNN architecture used for the 1D study}
\label{fig:CNN}
\end{figure}

After an iterative process of line search in the hyperparameters of the CNNs employed, it was determined that a typical optimal model for our phononic eigenvalue problem has the following architecture: There are 100 input nodes corresponding to the 100-element unit cell for the 1-D case (Fig. \ref{fig:CNN}). Each input node has two channels with individual channels corresponding to the density and modulus values of the element. This is a crucial piece of information in our CNN implementation in that we choose to identify the various material properties of an element with individual channels in a CNN. Not only does this create a direct correspondence with the practice of identifying the RGB information at a pixel with individual channels in computer vision applications, it also guides how CNNs could be employed for problems in mechanics in higher dimensions. In higher dimensions, more material properties (components of elasticity tensor) are required to describe the behavior of materials. Given the success of CNNs in our current 1-D problem, a promising strategy going forward would be to identify these material properties as different channels in a CNN input node layer. 

\begin{figure}[htp]
\centering 
\includegraphics[scale=0.77]{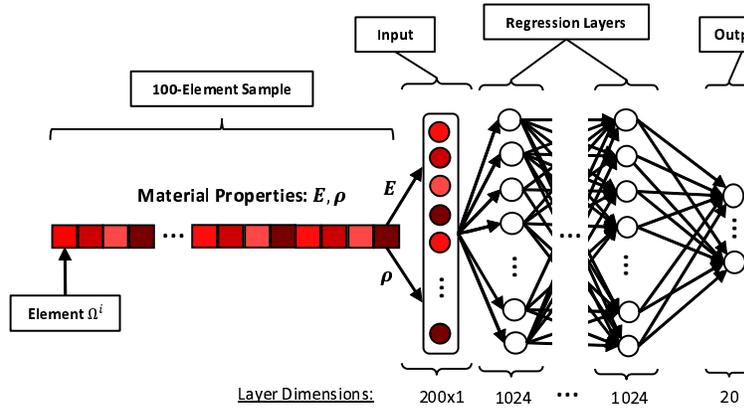}
\caption{MLP architecture used for the 1D study}
\label{fig:MLP}
\end{figure}

Following the input node is a set of two convolution layers with ReLU activation functions. The convolution filters has a size of 3 in the dimension of the input vector. A max-pooling layer follows the set of convolution layers with a filter dimension of 2$\times$1. These layers are followed by two fully connected dense layers with ReLU activation functions. Subsequently, a fully connected layer with linear or Gaussian connections leads to the outputs of the network which are the eigenvalues of the problem. For the purposes of most of the results shown below, the network was trained on 28,000 samples which took a little less than 13 minutes on a GPU (NVIDIA GTX980). Once the network is trained, prediction obviously takes a much smaller amount of time (fraction of a second.) The same input-output mapping was also performed using a multi-layer perceptron with only fully connected layers, which was used to generate some of the key results presented in this paper. The architecture that yielded the best prediction accuracy was found to have 6 hidden layers with 1024 computational units in each layer (Fig. \ref{fig:MLP}.) The final output layer consisted of gaussian or linear connections to the output units.    

\subsection{Results}

\subsubsection{Approximation Capabilities of CNNs}

Our initial focus was on obtaining average eigenvalue prediction accuracies higher than 95$\%$ on unseen examples and comparing the performance of CNNs vs regular MLPs in achieving this metric. We define mean absolute accuracy of our predictions:
\begin{equation}{}
e =  1-\frac{1}{p} \sum \frac{|(\hat{y} -y)|}{\hat{y}}
\end{equation}
where the summation is being carried out over all eigenvalues and all test data and $p$ is the total number of terms in the summation. The error is then expressed as a percentage. For the purpose of comparison, we considered Dataset-A for training, validation, and testing purposes. As mentioned earlier, Dataset-A has 300,000 data samples. A randomly selected portion of these is used for training the networks and the rest of the unseen examples are used for validation and testing purposes. 
\begin{figure}[htp]
\centering
\includegraphics[scale=0.51]{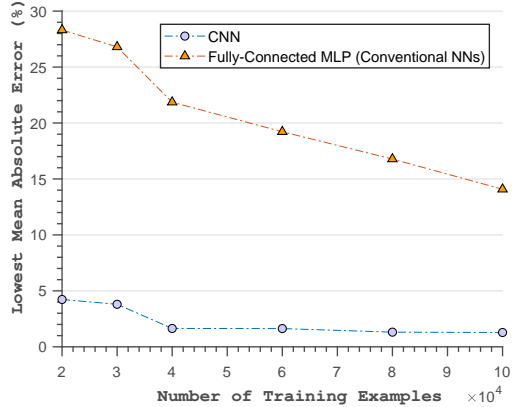} 
\caption{Comparison of prediction accuracy as a function of training data size (CNN vs MLP)}
\label{fig:MLPvsCNN}
\end{figure}

The striking point about this comparison is that the CNN easily outperforms the MLP in terms of eigenvalue prediction accuracy as can be seen in Fig. (\ref{fig:MLPvsCNN}). Both networks improve in their prediction accuracies as they are trained on larger fractions of Dataset-A. However, the CNN already has a higher than 95$\%$ prediction accuracy when it has been trained only on 20,000 samples. At this level of training, the MLP has an accuracy of slighly greater than 70$\%$. In fact, the MLP only reaches about 85$\%$ accuracy at 100,000 training samples at which point the CNN is already above 98$\%$ prediction accuracy. The 30,000-40,000 samples range seems to be a breaking point where both networks show the largest percental decrease of mean absolute error. In summary, these comparisons show that CNNs  have the potential to achieve high prediction accuracies in problems of mechanics with a fraction of the training data required by MLPs.

\begin{figure*}[htp]
    \centering
    \begin{subfigure}[t]{0.5\textwidth}
     \centering
     \includegraphics[scale=0.5]{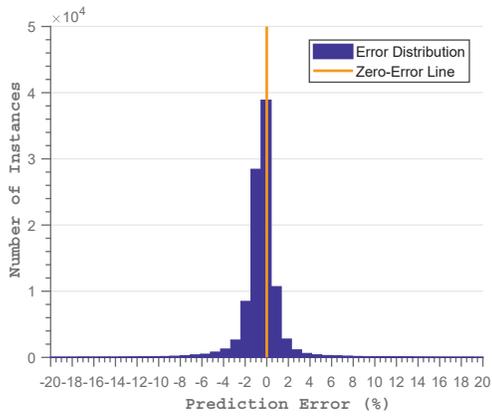} 
     \caption{Prediction error distribution}
    \end{subfigure}%
    ~ 
    \begin{subfigure}[t]{0.5\textwidth}
     \centering
     \includegraphics[scale=0.48]{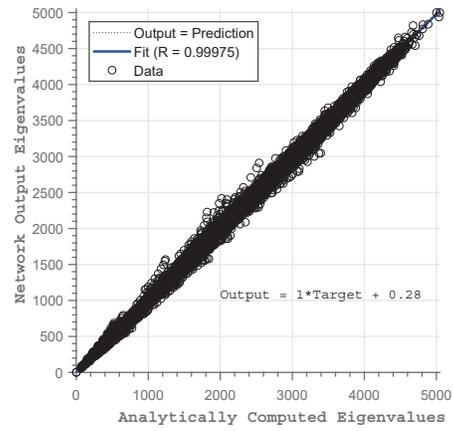} 
     \caption{Regression of the targets relative to the output}
    \end{subfigure}
    ~ 
	\begin{subfigure}[t]{0.5\textwidth}
     \centering
     \includegraphics[scale=0.5]{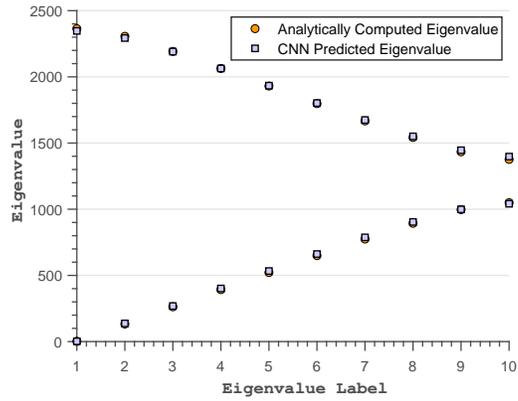} 
     \caption{Representative eigenvalue outputs for a 2-phase sample}
    \end{subfigure}
    \caption{CNN Performance}
    \label{fig:performance}
\end{figure*}

The CNN architecture employed was eventually able to achieve 1.13 $\%$ of mean absolute error in its eigenvalue predictions with a normalized 1-$\sigma$ error of  2.11e-05. The small value of the standard deviation shows that most of the distribution of prediction errors is heavily centered around its mean value. In other words, a large fraction of the predictions have absolute errors close to the reported mean of the distribution, whereas only a small fraction of the predictions has larger prediction errors. Fig. (\ref{fig:performance}b) shows all the predicted eigenvalues along with the associated analytically computed eigenvalues and clarifies the high correlation which exists between the two sets. This correlation can be measured in terms of the Pearson correlation coefficient which, for the present case, is at the level of 0.999 indicating that the regression between the two sets is strongly linear. This can also be seen from the distribution of prediction errors in Fig. (\ref{fig:performance}a) which shows that a vast majority of the predictions have errors in the $\pm 2\%$ range. Fig. (\ref{fig:performance}c) shows an example prediction of the CNN compared with actual eigenvalue calculations for a specific unseen phononic crystal configuration, which represents a sample from Dataset-2. Specifically, this sample represents a 2-phase material with two equal phases. One phase has a density of 5.80 g/cm$^3$ and a modulus of 76.27 GPa and the other phase has a density of 7.23 g/cm$^3$ and a modulus of 23.61 GPa. It can then be seen that all the 20 eigenvalues have been predicted with a reasonable accuracy by the CNN.

\subsubsection{Generalization Capabilities of CNNs}

\begin{figure*}[htp]
    \centering
    \begin{subfigure}[t]{0.5\textwidth}
     \centering
     \includegraphics[scale=0.51]{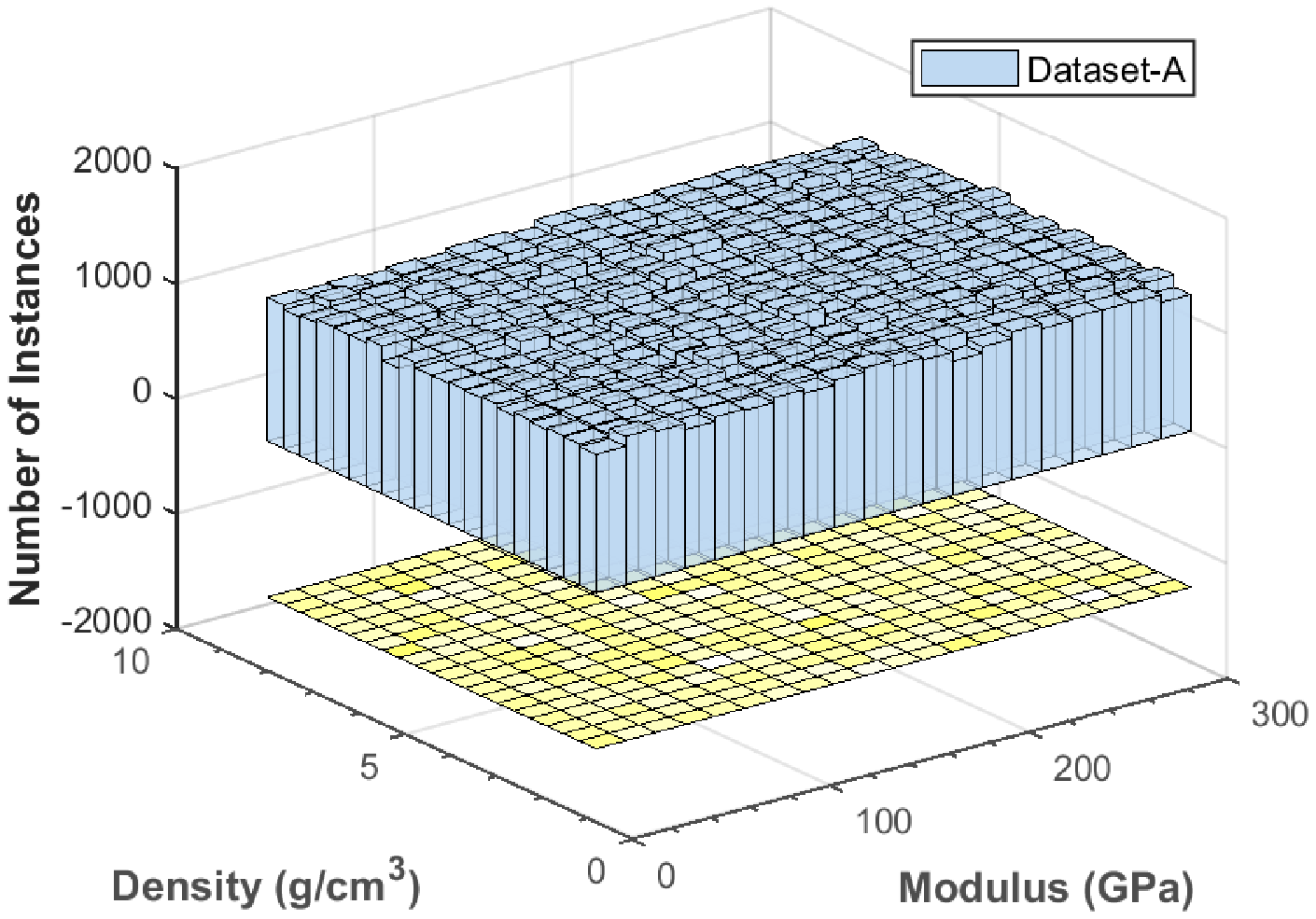} 
     \caption{Distribution of Input Material Properties for Dataset-A (100,000 samples)}
    \end{subfigure}%
    ~ 
    \begin{subfigure}[t]{0.5\textwidth}
     \centering
     \includegraphics[scale=0.51]{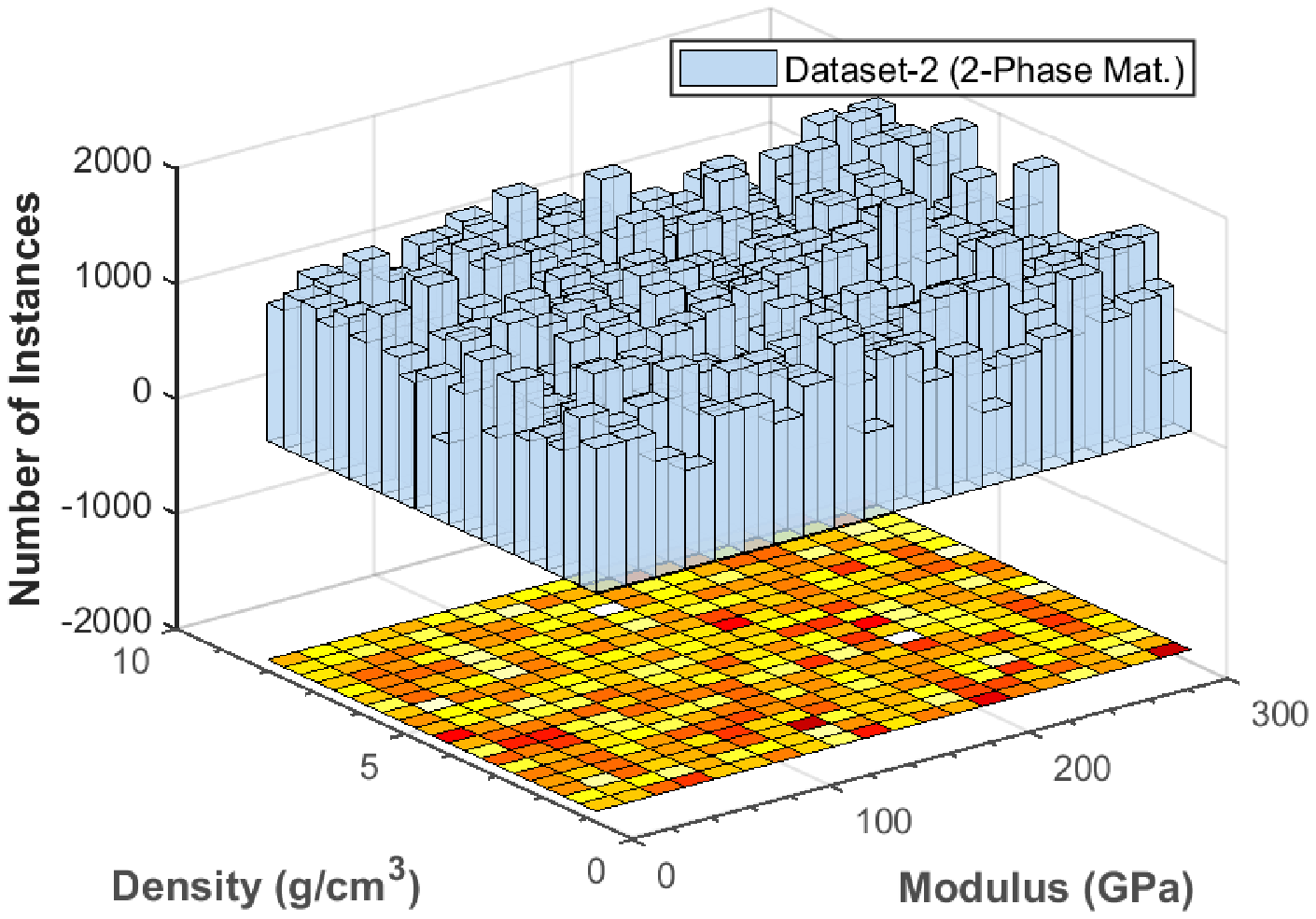} 
     \caption{Distribution of Input Material Properties for Dataset-2 (100,000 samples)}
    \end{subfigure}
    \caption{Comparison of the distribution on input material properties per element. Color map projection of the histograms is shown on the lower part, with the lightest color representing the most instances.}
    \label{fig:GeneralizationInput}
\end{figure*}
\begin{figure}[htp]
\centering
\includegraphics[scale=0.51]{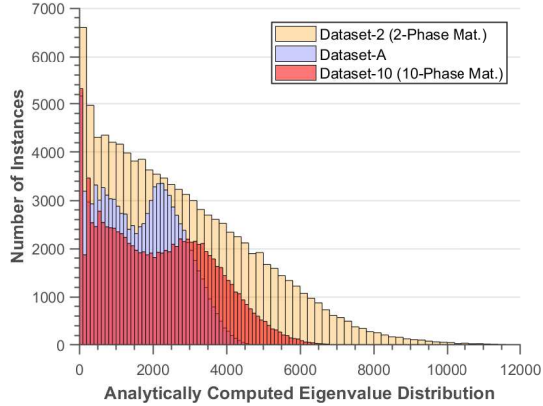} 
\caption{Uniform sample from the N100 and the 2 real materials datasets, regarding the 1st upper eigenvalue from the band structure}
\label{fig:GeneralizationOutput}
\end{figure}

From the previous section, it is evident that CNNs can massively outperform MLPs with fully connected layers in terms of their prediction accuracies for comparable training datasets. Another important consideration in evaluating the efficacy of any approximating method is its capability to generalize to unseen examples. To a limited extent, we have already shown that the trained CNN achieves very high prediction accuracies on unseen test data when the test data is extracted from the same distribution as the training and validation datasets. Here we consider the ability of the CNN to generalize to examples which are derived from significantly different distributions. For this we consider Dataset-A, Dataset-2, Dataset-3, and Dataset-10 as described earlier. While each of the elements in the samples in Dataset-A have materials properties that come from a random uniform distribution, the rest of the datasets used in this section contains contrasting phases that naturally yields a broader range of eigenvalue distributions. Representative 3D histograms for Dataset-A and Dataset-2 can be seen in Fig. (\ref{fig:GeneralizationInput}). The colormap projection in Fig. (\ref{fig:GeneralizationInput}a) shows that there is largely an even distribution of samples spanning the property ranges for Dataset-A. For dataset-2 (\ref{fig:GeneralizationInput}b), the distribution is less even and it misses some density ranges that are present in Dataset-A. Most importantly, however, since Dataset-B corresponds to only 2-phase cases, its corresponding eigenvalue ranges are very different from those of the samples in Dataset-A. This is clarified in Fig. (\ref{fig:GeneralizationOutput}) which shows histogram plots of all the eigenvalues for Dataset-A,  Dataset-2, and Dataset-10. It shows that the eigenvalue range spanned by Dataset-2 is about twice as large as that spanned by Dataset-A.
Fig. (\ref{fig:GeneralizationOutput}) shows some results which underline the generalization capabilities of the employed CNNs. First, when the network is only trained on Dataset-A, it is able to achieve $<16\%$ prediction errors on Dataset-2, $<15\%$ prediction errors on Dataset-3, and $<13\%$ prediction errors on Dataset-10 (blue bars.) This is interesting because not only does any input configuration in Dataset-2, Dataset-3, and Dataset-10 most likely does not exist in Dataset-A, there exist large eigenvalue ranges in these datasets which the trained network was never trained on and has never seen. These prediction errors come down substantially when Dataset-A is augmented with some samples from Dataset-2 and Dataset-3 for training purposes. The results are shown by orange bars in Fig. (\ref{fig:GeneralizationOutput}). In this case, the prediction errors for Dataset-2, Dataset-3, and Dataset-10 are all below $6\%$. Note that in this case, although the network was never trained on Dataset-10, it was able to generalize well when tested on this dataset. From these results, there is strong indication that deep CNNs applied as we present seem to be able to successfully generalize on completely new regions of the input-output space, improving upon the typical issues brought by implementations of shallow and less diverse networks in mechanics. These deep networks also seem to require much less data to achieve and surpass the performance of traditional MLPs.
\begin{figure}[htp]
\centering
\includegraphics[scale=0.51]{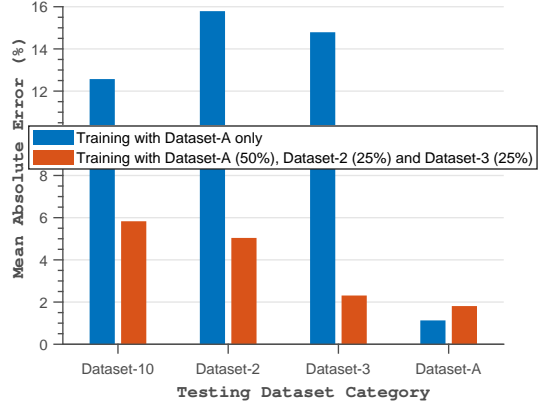} 
\caption{Model Generalization}
\label{fig:GeneralizationError}
\end{figure}

\subsubsection{Automatic learning of translational invariance}

An important and interesting feature of our CNN implementation is that it seems to have learned translational invariance of our problems automatically. Since phononic crystals are periodic composites, their eigenvalues (Eq. \ref{eq:Eigenvalue}) are invariant under unit cell translations. In effect, it means that a 2-phase unit cell with a 50-50 distribution of the two material phases ($P_1-P_2$) will have the same eigenvalues as a 3 phase unit cell made by periodically translating the two materials phases. For the latter, let's consider as an example, a unit cell made of phases $P_1-P_2-P_1$ with a 25-50-25 distribution respectively. Our CNN accurately predicts the eigenvalues for the two cases (both unseen), showing that it has learned the translation invariance of the problem. A deeper understanding of this can be done by reconstructing the activations of the deeper layers of the network and understanding the filter activations when a prediction is being made. For the most part, it is often not a trivial task to try to reconstruct the deeper layers of the network into the input space. However, a simple activation analysis on the maxpooling layer is performed in this study. The activations of the maxpooling filters feed directly to the regression layers. Given those activations are similar in the two cases, then the network prediction will also be the same. In theory and if the network's optimization is done successfully, filters should specialize such that their unique and sparse activation leads to the best output prediction possible. This means that input samples with similar input-output mappings would activate the same filters, while keeping the rest inactive.
\begin{figure}[htp]
\centering
\includegraphics[scale=0.51]
{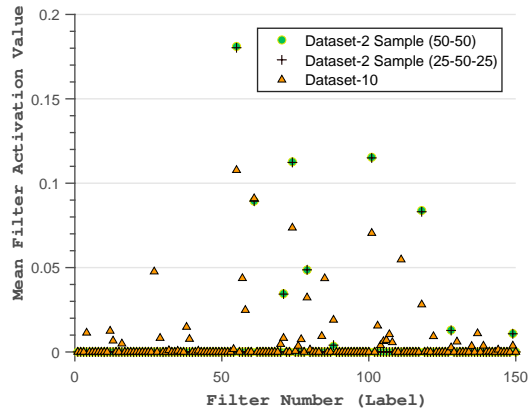}
\caption{Mean post-pooling filter activations versus filter number for two symmetric samples from Dataset-2 (which yields the same eigenvalue output) and a Dataset-10 sample}
\label{fig:activations1}
\end{figure}
Fig. (\ref{fig:activations1}) shows the maxpooling filter activation results when the network is presented with three different input unit cells. Two of these (from Dataset-2) are physically the same, differing only with a translation, with the third one being significantly different as it is taken from Dataset-10. We notice from this figure that for the cases of the translationally equivalent unit cells, not only are the same filters activated, but their activation values are also practically the same. On the other hand, for the sample from Dataset-10, the filter activations differ not only in their values, but also in which filters are activated.

\subsection{Convolutional Neural Networks for 2-D phononic eigenvalues}
\begin{figure}[htp]
\centering
\includegraphics[scale=0.25]{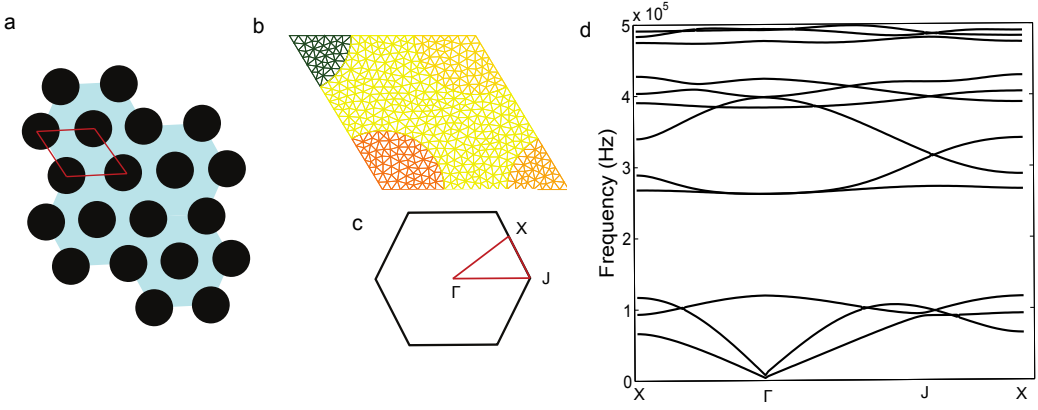}
\caption{a. Schematic of the 2-D periodic composite made from steel cylinders distributed in hexagonal packing in epoxy matrix,
b. Discretization of the unit cell, c. Irreducible Brouillon Zone in the reciprocal lattice, d. Band-structure calculation results
using the mixed variational formulation.}
\label{fig:fCompHexagonal}
\end{figure}
The formalism of input-output relationships as described in Eqs. (\ref{eq:1DIOC}, \ref{eq:1DIOD}) immediately applies to higher dimensional eigenvalue problems. Here, as a demonstration of the convolutional neural networks in higher dimensions, we consider the phononic eigenvalue problem emerging in a hexagonal phononic crystal. A representative example of 2-D phononic eigenvalues is shown in the bandstructure calculations given in Fig. (\ref{fig:fCompHexagonal}). More details about the calculations can be found in Ref. \cite{srivastava2014mixed}. In order to apply a CNN framework to the 2-D problem, we divide the hexagonal unit cell into a hexagonal grid of size ($128\times 128$). Now the eigenvalue input-output formalism is given by:
\begin{eqnarray}
\mathbf{e}=\mathbf{f}_d(\mathbf{E},\boldsymbol{\nu},\boldsymbol{\rho},Q_1,Q_2),\; \mathbf{E}=\{E_{ij}\},\boldsymbol{\nu}=\{\nu_{ij}\},\boldsymbol{\rho}=\{\rho_{ij}\};\;i,j=1,2...128
\label{eq:2DIOD}
\end{eqnarray}
where $E_{ij},\nu_{ij}$ are the value of the Young's modulus and Poisson's ratio respectively for the $i,j$ element in the hexagonal grid and $Q_1,Q_2$ are the normalized wavenumbers in the directions $\mathbf{q}^1,\mathbf{q}^2$ respectively. The training-testing-validation datasets are created by randomly generating material property matrices $\mathbf{E},\boldsymbol{\nu},\boldsymbol{\rho}$ by considering the Young's modulus as a random variable distributed between 100 MPa and 300 GPa, Poisson's ratio as a random variable betweeen 0.2 and 0.45, and density as a random variable between 800 and 8000 kg/m$^3$. As inputs to the neural networks, two additional matrices of size $128\times 128$  corresponding to $Q_1, Q_2$ are also considered. All the elements of one of these matrices are equal to $Q_1$ and all the elements of the other matrix are equal to $Q_2$. The matrices corresponding to $\mathbf{E},\boldsymbol{\nu},\boldsymbol{\rho},Q_1,Q_2$ are now concatenated into a 3-D matrix $\mathbf{I}$ of size $128\times 128\times 5$. The output $\mathbf{e}$ is a 1-D vector which consists of the lowest 50 eigenvalues of the phononic problem corresponding to $\mathbf{I}$. Four different datasets were created to demonstrate the training and testing of the neural network. Dataset-A consists of 100,000 samples created by considering the material properties in each element as independent random variables. Dataset-2, Dataset-3, and Dataset-5 each consist of 20,000 samples where, in each data-sample, the material properties are only allowed to take 2, 3, and 5 different randomly generated values respectively. 

\begin{figure*}[htp]
    \centering
    \begin{subfigure}[t]{0.5\textwidth}
     \centering
     \includegraphics[scale=0.51]{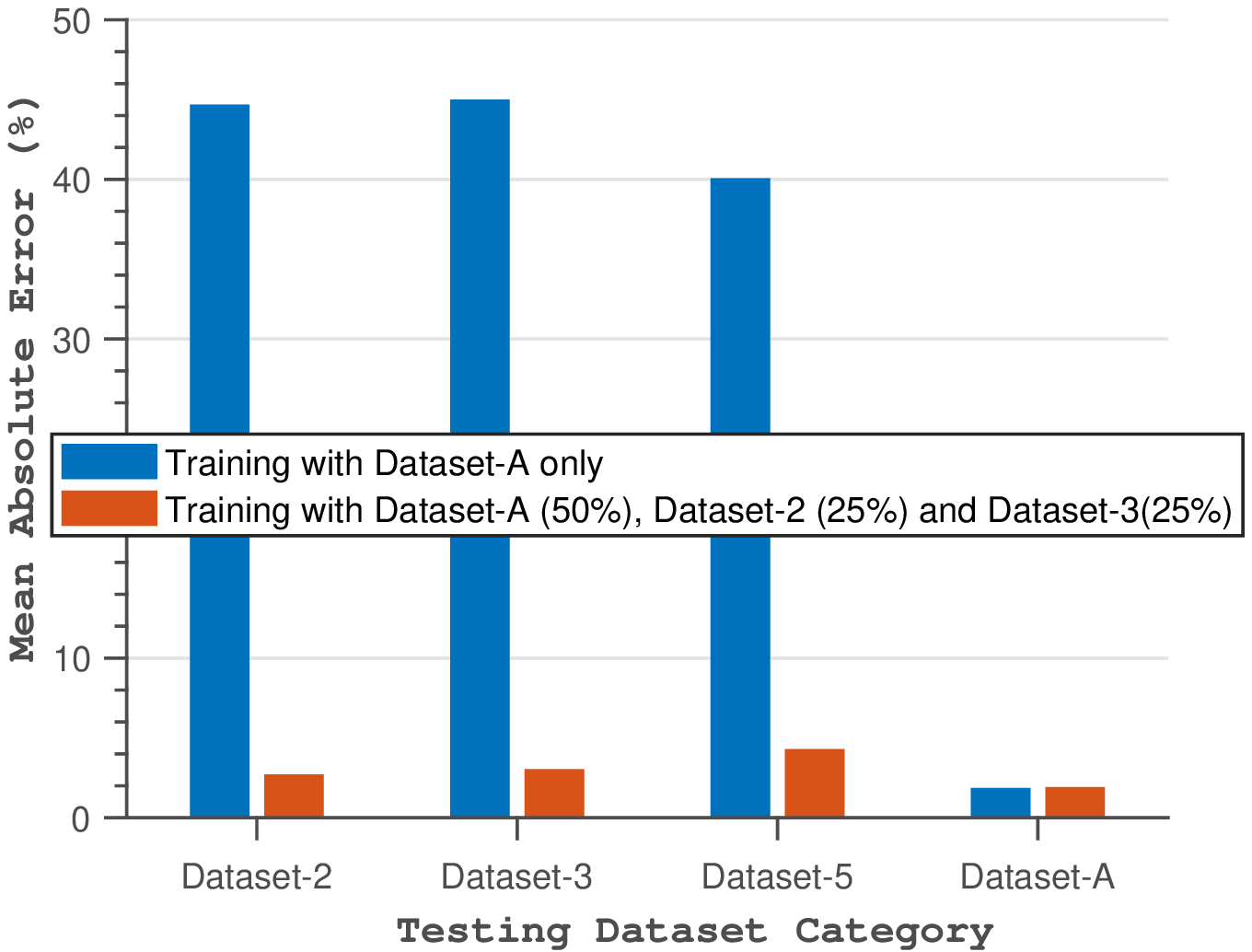} 
     \caption{Model Generalization for 2-D eigenvalue problem}
    \end{subfigure}%
    ~ 
    \begin{subfigure}[t]{0.5\textwidth}
     \centering
     \includegraphics[scale=0.51]{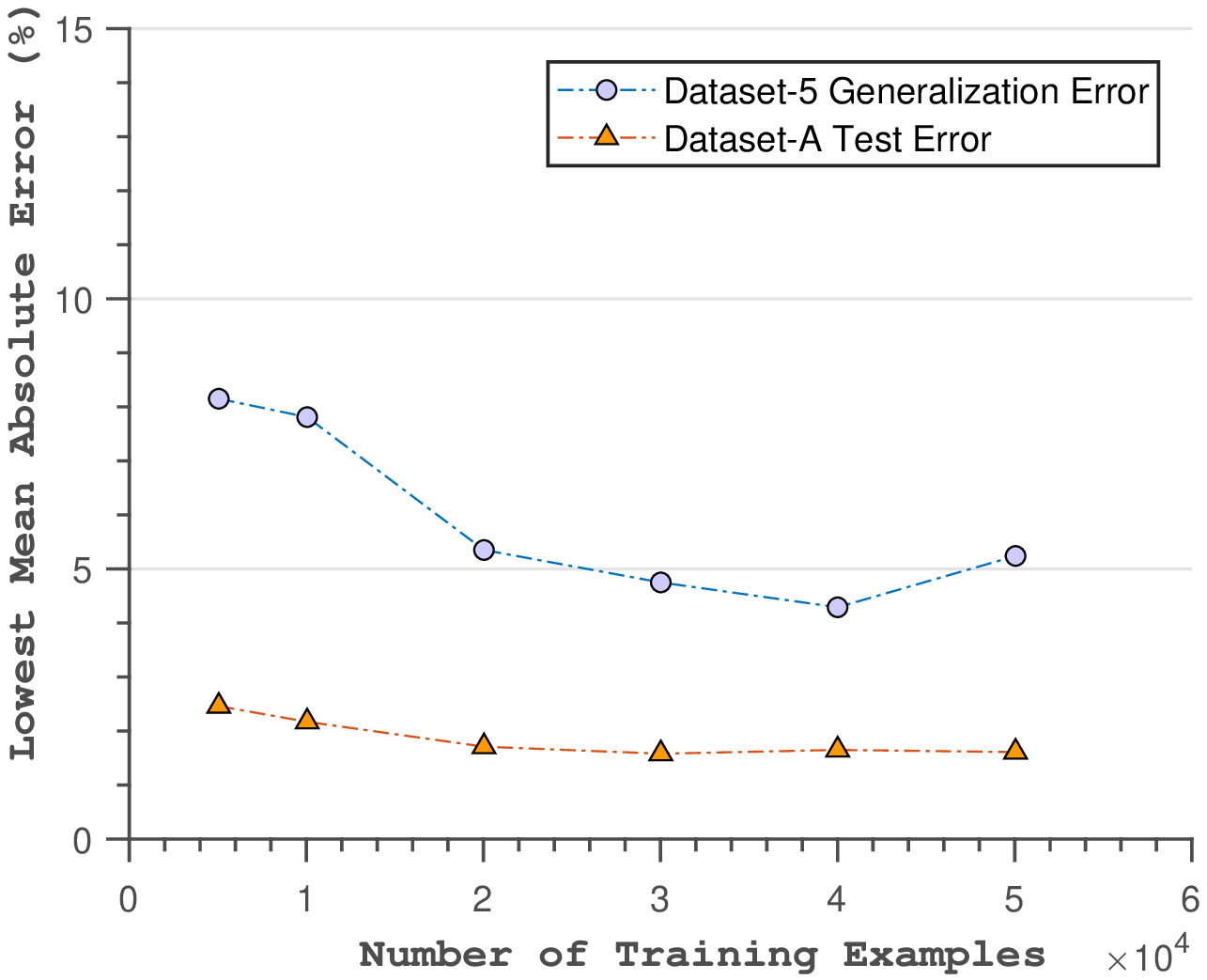} 
     \caption{Model test and generalization performance as a function of training data size}
    \end{subfigure}
    \caption{CNN prediction results on the 2-D eigenvalue problem}
    \label{fig:2DResults}
\end{figure*}

The optimized network architecture used for this problem is a natural extension of the one used in the 1-D problem.  The input space $\mathbf{I}$, with depth 5 in this case, is processed by 2D convolution and pooling filters, with size of 2x2 instead. These then lead to fully-connected units, which eventually connect to the output vector $\mathbf{e}$. Fig. (\ref{fig:2DResults}) shows some results which underline the generalization capabilities of the employed CNNs for the 2-D case. First, when the network is only trained on Dataset-A, it is able to achieve prediction errors below $50\%$ prediction errors on Dataset-2, Dataset-3, and Dataset-5 (blue bars). This difference with respect to the 1D case is likely tied to the significant increase on the input dimensionality. A significant decrease on prediction error is observed when the optimization dataset includes Dataset-A, Dataset-2 and Dataset-3 samples. The results are shown by orange bars in Fig. (\ref{fig:2DResults}a).  As seen in the orange bars in Fig. (\ref{fig:2DResults}a), the prediction errors for Dataset-2, Dataset-3, and Dataset-5 are all below $5\%$. Just as in the 1D case, the network was never trained on Dataset-5, but it was able to also generate predictions with errors below the target threshold in this new space. Finally, Figure (\ref{fig:2DResults}b) shows how the prediction error on unseen Dataset-5 samples changes as the number of total samples in the optimized mixed dataset increases. This demonstrates the data efficiency and accuracy of the CNN implementation stands for this higher dimensional case as well.

\section{Conclusion}

We have shown that key mathematical properties that have made modern deep neural network architectures, such as CNNs, and that have proven to be very successful in typical machine learning implementation fields, can translate to complex input-output mapping tasks in mechanics. Through our results, not only were we able to demonstrate that CNNs can successfully approximate the complex function mapping of material tensors of a phononic crystal to its eigenvalues, but also that it can be done with much higher data efficiency and performance than traditional neural networks (densely connected MLPs). The CNN model architectures used in this study present a much more diverse set of computational elements and deeper graph depth that aid both the feature abstraction process and the function approximation capabilities, as is well established in the machine learning literature. The basic filter activation study in this paper also provided a hint at how the feature abstraction and function mapping occurs in our implementation, leaving the door open for more a comprehensive study in future applications. We found that when using an optimized network to predict a case with a characteristic input translation that leads to the same output, exactly the same post-pooling filters are consistently activated with quite similar activation values. This fact provides clues that these filters are specializing in characteristic input patterns that aids the prediction of the output eigenvalues. This process has been noted to occur in computer vision applications of deep CNNs and have proven useful in our application in mechanics as well. 

There are many interesting questions worth pursuing after this study. The primary one is the extension of this process to dimensions higher than 2 for the prediction of eigenvalues in more complex mechanics problems. In higher dimensions more material properties would need to be taken into account. For instance, even if we consider an isotropic material, its mechanical description would generally require 3 independent constants (shear modulus, bulk modulus, and density). For anisotropic materials this number further goes up. However, within the CNN framework presented in this paper, taking these into account merely requires the addition of more channels or depth to the input vectors. With this framework, it would be interesting to see if deep CNNs can be efficiently trained to act as substitutes for eigenvalue algorithms in the case of 2-, and 3-D mechanics problems. Such networks would have no fundamental computational complexity limitations with which all eigenvalue algorithms suffer and, therefore, would provide a way to explore design spaces which have not been probed yet. The idea of representing the elements of material property tensors as different channels in a CNN, as proposed in this paper, is clearly not limited to eigenvalue problems. Therefore, it is conceivable that CNNs can similarly lead to significant improvements in regression and classification tasks in non-eigenvalue mechanics problems including time domain problems.


\section*{Acknowledgements}
A.S. acknowledges support from the NSF CAREER grant \# 1554033 to the Illinois Institute of Technology.

\section*{References}
\bibliography{ReferencesBib}

\end{document}